\documentclass[twocolumn]{aastex63}
\usepackage{amssymb,amsmath,bm}
\usepackage{graphicx}
\usepackage{natbib}
\usepackage{hyperref}

\received{xxx xx, 2023}
\revised{xxx xx, 2023}
\accepted{xxx xx, 2023}
\submitjournal{xxx}

\begin{document}
\title{Cosmological Structure Formation and Soliton Phase Transition in Fuzzy Dark Matter with Axion Self-Interactions}

\correspondingauthor{Philip Mocz}
\email{mocz1@llnl.gov}

\author[0000-0001-6631-2566]{Philip Mocz}
\affiliation{Department of Astrophysical Sciences, Princeton University, 4 Ivy Lane, 
Princeton, NJ 08544, USA}
\affiliation{Lawrence Livermore National Laboratory, 4 Ivy Lane, 7000 East Ave, Livermore, CA 94550, USA}
\author[0000-0002-1369-633X]{Anastasia Fialkov}
\affiliation{Institute of Astronomy, University of Cambridge, Madingley Road, Cambridge CB3 0HA, UK}
\affiliation{Kavli Institute for Cosmology, Madingley Road, Cambridge CB3 0HA, UK}
\author[0000-0001-8593-7692]{Mark Vogelsberger}
\affiliation{Department of Physics, Kavli Institute for Astrophysics and Space Research, Massachusetts Institute of Technology, Cambridge, MA 02139, USA}
\author[0000-0002-9604-343X]{Michael Boylan-Kolchin}
\affiliation{Department of Astronomy, The University of Texas at Austin, 2515 Speedway, Stop C1400, Austin, TX 78712-1205, USA}
\author[0000-0002-2386-3904]{Pierre-Henri Chavanis}
\affiliation{Laboratoire de Physique Th\'eorique, Universit\'e Paul Sabatier, 118 route de Narbonne 31062 Toulouse, France}
\author[0000-0002-8742-197X]{Mustafa A. Amin}
\affiliation{Physics \& Astronomy Department, Rice University, Houston, Texas 77005-1827, USA}
\author[0000-0002-0974-5266]{Sownak Bose}
\affiliation{Institute for Computational Cosmology, Department of Physics, Durham University, Durham, DH1 3LE, UK}
\author[0000-0003-2586-3702]{Tibor Dome}
\affiliation{Institute of Astronomy, University of Cambridge, Madingley Road, Cambridge, CB3 0HA, UK}
\affiliation{Kavli Institute for Cosmology, Madingley Road, Cambridge CB3 0HA, UK}
\author[0000-0001-6950-1629]{Lars Hernquist}
\affiliation{Harvard-Smithsonian Center for Astrophysics, 60 Garden Street, Cambridge, MA 02138, USA}
\author[0000-0002-0041-4356]{Lachlan Lancaster}
\affiliation{Department of Astronomy, Columbia University,  550 W 120th St, New York, NY 10025, USA}
\affiliation{Center for Computational Astrophysics, Flatiron Institute, 162 5th Avenue, New York, NY 10010, USA}
\author[0000-0002-5030-3718]{Matthew Notis}
\affiliation{Department of Astrophysical Sciences, Princeton University, 4 Ivy Lane, 
Princeton, NJ 08544, USA}
\author[0000-0002-3531-4806]{Connor Painter}
\affiliation{Department of Astronomy, The University of Texas at Austin, 2515 Speedway, Stop C1400, Austin, TX 78712-1205, USA}
\author[0000-0002-9497-9963]{Victor H. Robles} 
\affiliation{Physics Department, Yale Center for Astronomy and Astrophysics, New Haven, CT 06520, USA}
\author[0000-0003-4442-908X]{Jes\'us Zavala}
\affiliation{Center for Astrophysics and Cosmology, Science Institute, University of Iceland, Dunhagi 5, 107 Reykjavik, Iceland}


\begin{abstract}
We investigate cosmological structure formation in Fuzzy Dark Matter (FDM) with an attractive self-interaction (SI) with numerical simulations. Such a SI would arise if the FDM boson were an ultra-light axion, which has a strong CP symmetry-breaking scale (decay constant). Although weak, the attractive SI may be strong enough to counteract the quantum `pressure' and alter structure formation. We find in our simulations that the SI can enhance small-scale structure formation, and soliton cores above a critical mass undergo a phase transition, transforming from dilute to dense solitons.
\end{abstract}


\section{Introduction}
\label{sec:intro}

Ultra-light bosons continue to be a popular candidate for the dark matter in our Universe \citep{2000PhRvL..85.1158H,2003PhRvD..68b4023G,2017PhRvD..95d3541H,2019PhRvL.123n1301M,2020ApJ...904..161B,2020PrPNP.11303787N,2021arXiv210111735H}.
The so-called Fuzzy Dark Matter (FDM) model postulates a particle mass of $m\sim10^{-22}~{\rm eV}$, 
which introduces wave dynamics in the dark matter on the de Broglie wavelength $\lambda_{\rm dB}\sim 1~{\rm kpc}$ -- the scale of galaxies.
The arising quantum `pressure' (really a pressure tensor) suppresses small-scale power in the initial dark matter power spectrum \citep{2000PhRvL..85.1158H},
modifies the halo mass function 
\citep{2020PhRvD.101l3026S}, 
and creates \textit{soliton} cores at the centers of dark matter halos
\citep{2014NatPh..10..496S,2015MNRAS.451.2479M}.
Solitons are quasi-stable cored objects with total mass scaling inversely with radius, unique to the FDM model.
FDM has seen a rise in direct numerical simulations that investigate non-linear and small-scale features of the model \citep{2014NatPh..10..496S,2017MNRAS.471.4559M,2019PhRvL.123n1301M,2020MNRAS.494.2027M,2018PhRvD..97f3507D,2020PhRvD.101h3518V,2020PhRvD.102h3518S,2020arXiv200408482L,2021MNRAS.501.1539N,2021PhRvD.103b3508L,2021arXiv210101828M}.
A challenge for the FDM model continues to be to understand whether an ultra-light particle mass can simultaneously predict the Lyman-$\alpha$ forest power spectrum extracted from high-redshift quasars  
\citep{2017PhRvL.119c1302I,2019MNRAS.482.3227N}
as well as explain the core sizes of satellite galaxies
\citep{2020ApJ...893...21S,2020ApJ...904..161B}.
This is a \textit{Catch-22} problem \citep{2020MNRAS.492.5721D} of sorts, in that smaller boson masses lead to larger, less dense cores, but also less structure in the Lyman-$\alpha$ forest.
\cite{dalal2022not} use sizes and stellar kinematics of ultra-faint dwarf galaxies to place a strict lower-limit of $m>3\cdot10^{-19}~{\rm eV}$ in the simple FDM model, arguing the 1-parameter family of soliton solutions at the center of halos cannot fit the observational data at lower particles masses (neglecting dynamical heating).

This Catch-22 may be resolved by the introduction of a second relevant scale, determined by the way that FDM particles interact with one another. This could arise naturally in one of the main candidate models for the FDM boson: the hypothetical axion arising from the symmetry breaking needed to solve the strong CP Problem \citep{1977PhRvL..38.1440P,1978PhRvL..40..223W}. In this model, the axion would have a decay constant (or symmetry-breaking scale) $f$ associated with it, 
which would give rise to an attractive self-interaction (SI) \citep{2018PhRvD..97b3529D}.
Such an ultra-light axion may constitute a considerable fraction of the present-day critical density of the Universe (e.g. \citealt{2016PhR...643....1M,2017PhRvD..95d3541H,2018PhRvD..97b3529D}):
\begin{equation}
\label{eqn:abundance}
    \Omega \sim 0.1 \left(\frac{f}{10^{17}~{\rm GeV}}\right)^2 \left(\frac{m}{10^{-22}~{\rm eV}}\right)^{1/2}.
\end{equation}
For fiducial values $m\sim10^{-22}~{\rm eV}$ and $f\sim 10^{17}~{\rm GeV}$, 
the  attractive SI is \textit{tiny}: the dimensionless strength of the quartic coupling
$m^2/f^2\sim10^{-96}$, and hence the attractive SI has thus far been ignored in most numerical simulations.
Despite this tiny value,
the analytical findings of
\cite{2018PhRvD..97b3529D}
indicate that the cosmic web is
influenced by a small, non-vanishing self-coupling among
ultra-light axions.
\cite{2018PhRvD..97b3529D} show that attractive SI can have a significant impact on the stability of cosmic structures at low
redshift, including filaments and soliton cores.
A noticeable effect on cosmological scales is likely seen at $f\lesssim 10^{13}~{\rm GeV}$.
Other analytic studies have also indicated that attractive SI would only allow solitons to remain stable below some critical maximum mass \citep{1973R&QE...16..783V,2011PhRvD..84d3531C,2016PhRvD..94h3007C}. Below that mass, solitons are in a {\it dilute} phase \citep{2011PhRvD..84d3532C}.  Above that mass, the solitons collapse and form {\it dense} solitons \citep{2016PhRvL.117l1801B} which are stabilized by higher-order repulsive terms in the expansion of the self-interaction potential \citep{2016JHEP...12..066E,2018PhRvD..98b3009C}.\footnote{Some authors \citep{2018PhLB..777...64V,2019PhRvD.100f3002E} argue that, when relativistic effects are taken into account,  dense solitons
made of a real axionic SF are unstable and decay via emission of relativistic axions on a timescale much
shorter than any cosmological timescale. This conclusion is, however, not universally accepted \citep{2019RvMP...91d1002B}.} The collapse of the solitons may be accompanied by a sort of ``explosion'' (a burst of relativistic axions) leading to a bosenova \citep{2017PhRvL.118a1301L}. The bosenova phenomenon occurs in the case of a relatively strong self-interaction $f< M_P\sim 10^{19}\, {\rm GeV}$. In certain regimes, not relevant here, it is necessary to take general relativity into account and the collapse rather leads to a black hole \citep{2017JCAP...03..055H}.

The goal of this paper is to offer the first cosmological simulation of FDM with attractive SI and study the impact of instabilities on structure formation in the post-recombination universe.  Local numerical simulations with attractive SI have seen performed recently \citep{2021PhRvD.104h3022C,2021PhRvD.104h3532G} at the scale of one cluster in a static background. 
Cosmological simulations including gravity and attractive self-interactions in an expanding universe were carried out in \cite{Amin:2019ums} using the Schr\"{o}dinger-Poisson system. In that work, however, the focus was on soliton formation and their gravitational clustering rather than late-time structure formation.\footnote{In terms of their fiducial parameters, a much stronger self-interaction strength was used than the one considered here.}

Our manuscript is organized as follows.
In \S~\ref{sec:setup} we lay out the non-relativistic limit for the axion dark matter model, relevant for our cosmological simulations.
In \S~\ref{sec:sim} we describe the simulations.
In \S~\ref{sec:ps} we discuss the impact of SI on the dark matter power spectrum.
In \S~\ref{sec:soliton} we explore the phase transition that affects dark matter  solitons due to the SI.
We offer our concluding remarks in \S~\ref{sec:conc}.

\begin{table*}[ht]
    \centering
    \begin{tabular}{cccccccccc}
    \hline
    sim. & DM & $m~[{\rm eV}]$ & $a_s~[{\rm cm}]$ & $f~[{\rm GeV}]$ & $M_{\rm max}~[M_\odot]$ & $M_{\rm min}~[M_\odot]$ & $M_{1/2}~[M_\odot]$ & res. & $L_{\rm box}~[h^{-1}{\rm Mpc}]$ \\ 
    \hline
    1 & CDM   & --         & -- & -- & -- & -- & -- &  $512^3$ & $1.5$ \\
    2 & FDM   & $10^{-22}$ & -- & -- & -- & $1.4\times 10^{7}$ & $5\times 10^{10}$ & $1024^3$ & $1.5$ \\
    3 & SIFDM & $10^{-22}$ & $-1\times 10^{-75}$ & $1.4\times 10^{14}$ & $1.6\times10^8$ & $1.4\times 10^{7}$ & $5\times 10^{10}$ &  $1024^3$ & $1.5$ \\
    4 & SIFDM & $10^{-22}$ & $-4\times 10^{-75}$ & $7.0\times 10^{13}$ & $7.8\times10^7$ & $1.4\times 10^{7}$ & $5\times 10^{10}$ &  $1024^3$ & $1.5$ \\
    5 & SIFDM & $10^{-22}$ & $-8\times 10^{-75}$ & $5.0\times 10^{13}$ & $5.5\times10^7$ & $1.4\times 10^{7}$ & $5\times 10^{10}$ &  $1024^3$ & $1.5$ \\
    \hline
    \end{tabular}
    \caption{Simulation parameters and setup.
    $m$ is the axion mass, 
    $a_s$ is the effective $s$-scattering length of the self-interaction, which can equivalently be defined by the  axion decay constant $f$.
    $M_{\rm max}$ is the maximum stable soliton mass.
    $M_{\rm min}$ is the minimum mass halo formed in the cosmological simulation.
    $M_{1/2}$ is the cutoff scale in the initial cosmological power spectrum from linear theory.
    }
    \label{tab:sims}
\end{table*}

\section{FDM with Attractive Self-Interaction}
\label{sec:setup}

We assume a real scalar field $\phi$ 
in the weak-field limit in an expanding universe,
with an instantonic axion potential \citep{1977PhRvL..38.1440P,1980AnPhy.128..363W,1980NuPhB.171..253D} $\mathcal{V}(\phi)$:
\begin{equation}
\label{eqn:instanton}
\mathcal{V}(\phi) = \frac{m^2cf^2}{\hbar^3}
\left( 1-\cos\left(\frac{\hbar^{1/2}c^{1/2}\phi}{f}\right) \right) - \frac{m^2c^2}{2\hbar^2}\phi^2.
\end{equation}
Such a system is governed by the  Klein-Gordon-Einstein (KGE) equations.

In the non-relativistic limit ($c\to\infty$), making the Klein transformation
\begin{equation}
\phi = \frac{1}{\sqrt{2}}
\frac{\hbar}{m}
\left(
\psi(\mathbf{x},t)e^{-imc^2t/\hbar}
+
\psi^*(\mathbf{x},t)e^{imc^2t/\hbar}
\right)
\end{equation}
to separate the fast oscillations (with a pulsation $\omega=mc^2/\hbar\gg H$) from the slow evolution of the complex wavefunction $\psi$, the KGE equations reduce to the 
Gross-Pitaevski-Poisson (GPP) equations
in an expanding universe:
\begin{eqnarray}
\label{eqn:gp}
i\hbar \left(\frac{\partial }{\partial t} +\frac{3}{2}H \right) \psi &=& -\frac{\hbar^2}{2m}\nabla^2\psi + m V\psi  \nonumber \\
&-&\frac{4\pi\hbar^2 |a_s|}{m^2}\lvert\psi\rvert^2\psi 
+\frac{32\pi\hbar^4 |a_s|^2}{3m^5c^2}\lvert\psi\rvert^4\psi,\nonumber\\
\end{eqnarray}
\begin{equation}
\label{eqn:p}
\nabla^2 V = 4\pi G(\rho-\overline{\rho}), 
\end{equation}
where $H$ is the Hubble constant, 
$V$ is the gravitational potential 
seeded by the density $\rho\equiv |\psi|^2$,
and $a_s<0$ is an effective $s$-scattering length 
of the SI related to the axion decay constant $f$ via:
\begin{equation}
f = \sqrt{\frac{\hbar c^3 m}{ 32 \pi |a_s|}}.
\end{equation}
For a detailed derivation of Eqns.~\eqref{eqn:gp} and \eqref{eqn:p}, see \cite{2018PhRvD..98b3009C}.
In the above, the Hubble constant $H\equiv \dot{a}/a$
encodes cosmological expansion, 
where $a\equiv 1/(1+z)$ is the cosmological expansion factor and
$z$ is the redshift.

Equations~\eqref{eqn:gp} and~\eqref{eqn:p} with $a_s=0$
are the Schr\"odinger-Poisson (SP) equations
(e.g. \citealt{2014NatPh..10..496S,2018PhRvD..97h3519M}) commonly
used to simulate FDM neglecting SI.
The $|\psi|^2$ and $|\psi|^4$ terms in the equation
come from a Taylor expansion of the non-relativistic limit of the instantonic axion potential Eqn.~\eqref{eqn:instanton}.
The $|\psi|^2$ is an attractive SI term.
The next-order $|\psi|^4$ term, only relevant at very high densities, 
is repulsive.

\subsection{Soliton Instability}

The SP equations admit a well-known stable ground state soliton solution, approximated analytically by \citep{2014NatPh..10..496S}:\footnote{The soliton can also be conveniently approximated by a Gaussian profile (see Fig. 2 in \cite{2019PhRvD.100h3022C}).}
\begin{equation}
\label{eqn:soliton}
\rho(r)\simeq
\rho_0
\left[1+0.091\times \left(\frac{r}{r_{\rm c}}\right)^2\right]^{-8},
\end{equation}
where $r$ is the spherical coordinate, $r_{\rm c}$ is the core radius, and $\rho_0$ is the central density:
\begin{equation}
\rho_0\simeq 1.9\times 10^{9} 
\left(\frac{10^{-22}~{\rm eV}}{m}\right)^2
\left(\frac{{\rm kpc}}{r_{\rm c}}\right)^4
\frac{M_\odot}{{\rm kpc}^3}.
\end{equation}
The soliton core has total mass $M_{\rm c}$:
\begin{equation}
M_{\rm c} \simeq 2.2\times10^{10}
\left(\frac{10^{-22}~{\rm eV}}{m}\right)^2
\left(\frac{{\rm kpc}}{r_{\rm c}}\right)
M_\odot.
\end{equation}

With attractive SI added, the soliton becomes unstable above a
maximum critical mass \citep{2011PhRvD..84d3531C,2018PhRvD..98b3009C}:
\begin{equation}
\label{eqn:max}
M_{\rm max} = 1.012\frac{\hbar}{\sqrt{Gm |a_s|}}
\end{equation}
triggering a \textit{phase transition} 
between dilute (Eqn.~\ref{eqn:soliton})
and dense solitons.
The precise outcome of a dense soliton
requires reverting back to the relativistic 
version of the governing physical equations; 
however, in the non-relativistic version of the equations,
the repulsive $|\psi|^4$ term may regularize and balance
the attractive $|\psi|^2$ term and form a compact object of
approximately constant density \citep{2016PhRvL.117l1801B,2018PhRvD..98b3009C}
\begin{equation}
\label{eqn:compact}
\rho_{\rm dense} \simeq \frac{9m^3c^2}{32\pi|a_s|\hbar^2}.
\end{equation}

\subsection{Linear Instability Scales}

For non-relativistic self-gravitating Bose-Einstein Condensates (BECs) with an attractive self-interaction in an expanding universe, the equation for the density
contrast in the linear regime is \citep{2012A&A...537A.127C}:
\begin{equation}
\label{eqn:perturb}
\ddot{\delta} + 2\frac{\dot{a}}{a} \dot{\delta} 
+ \left(
\frac{\hbar^2 k^4}{4m^2a^4}
-\frac{4\pi |a_s|\hbar^2\rho k^2}{m^3a^2}
-4\pi G \rho
\right) \delta = 0,
\end{equation}
where $\delta$ is the over-density parameter.

Equation~\eqref{eqn:perturb} can be obtained from the hydrodynamic representation of the GPP equations. Structure formation results from
the competition between the quantum pressure, the attractive self-interaction and the self-gravity. The competition
between the quantum pressure and the self-gravity defines a (comoving) quantum Jeans wavenumber \citep{1985MNRAS.215..575K}:
\begin{equation}
    k_J = \left( \frac{16\pi G\rho m^2a^4}{\hbar^2} \right)^{1/4}.
\end{equation}
The competition between the quantum pressure and the self-interaction defines a (comoving) self-interaction wavenumber \citep{2011PhRvD..84d3531C}:
\begin{equation}
\label{eqn:kI}
    k_I = \left( \frac{16\pi |a_s|\rho a^2}{m} \right)^{1/2}.
\end{equation}
When all effects (gravity, quantum pressure and self-interaction) are taken into account, the critical wavenumber
is obtained by putting the term in parenthesis in Eqn.~\eqref{eqn:perturb} equal to zero. This condition can be written as:
\begin{equation}
    k^4 - k_I^2 k^2 - k_J^4 = 0.
\end{equation}
Therefore, the critical wavenumber can be expressed in terms of $k_I$ and $k_J$ as:
\begin{equation}
\label{eqn:kc}
k_c^2=\frac{1}{2}\left( k_I^2 + \sqrt{k_I^4+4k_J^4} \right).
\end{equation}
Jeans-type instability occurs for $k < k_c$. 

In a cosmological context,
with density $\rho\propto a^{-3}$,
and a Hubble parameter $h=0.7$,
\cite{2018PhRvD..97b3529D}
rewrites the comoving instability scales as:
\begin{equation}
\frac{k_J(a)}{h\,{\rm Mpc}^{-1}} = 161\, a^{1/4} 
\left(\frac{m}{10^{-22}~{\rm eV}}\right)^{1/2}
\left(\Omega_m h^2 \right)^{1/4},
\end{equation}
\begin{equation}
\frac{k_I(a)}{h\,{\rm Mpc}^{-1}} = 0.015\, a^{-1/2}
\left(\frac{f}{10^{17}~{\rm GeV}}\right)^{-1}
\left(\Omega_m h^2 \right)^{1/2}\,,
\end{equation}
which allows us see their fiducial values and scaling.
These scales imply that in the simple FDM model structure formation happens at physical scales larger than the Jeans scale $k<k_J$ and with self-interaction, there is a secondary instability mode at $k<k_I$.

\begin{figure*}[ht!]
\centering
\includegraphics[width=\textwidth]{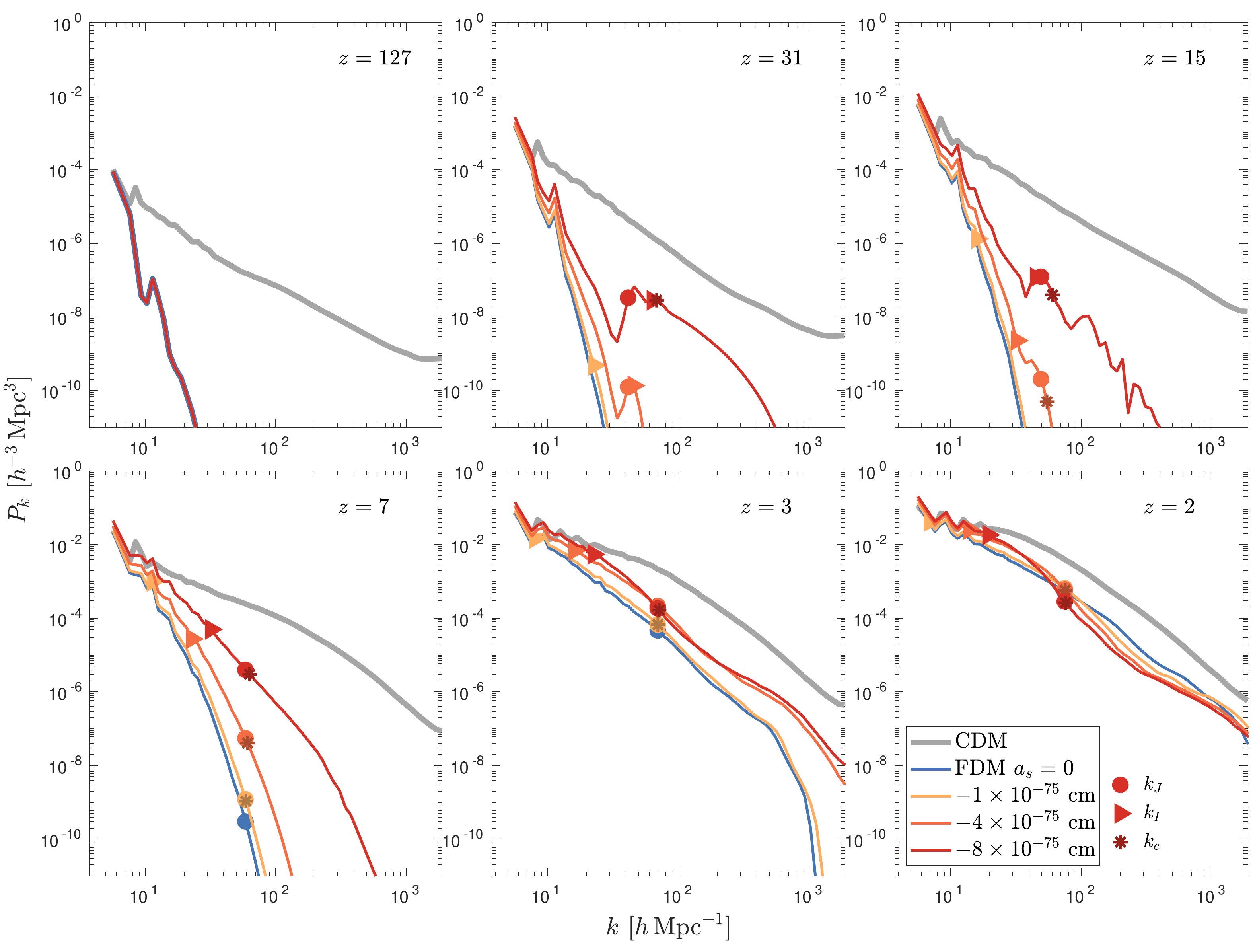}\\
\includegraphics[width=\textwidth]{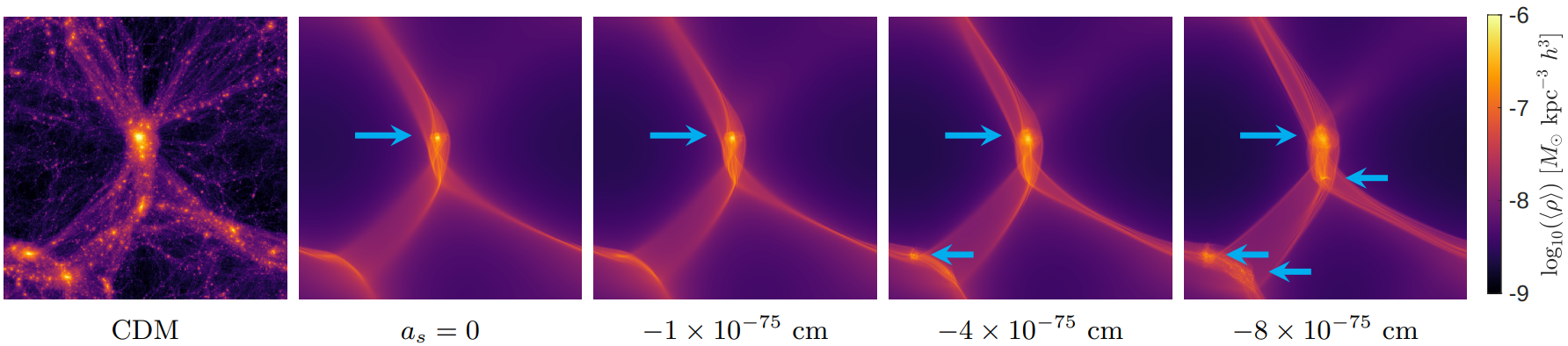}\\
\caption{Evolving dark matter power spectra of our numerical simulations, with instability scales indicated.
Shown also on the bottom row are projected dark matter densities at $z=2$, with blue arrows denoting formed solitons.}
\label{fig:main}
\end{figure*}

\begin{figure}[ht!]
\centering
\includegraphics[width=0.47\textwidth]{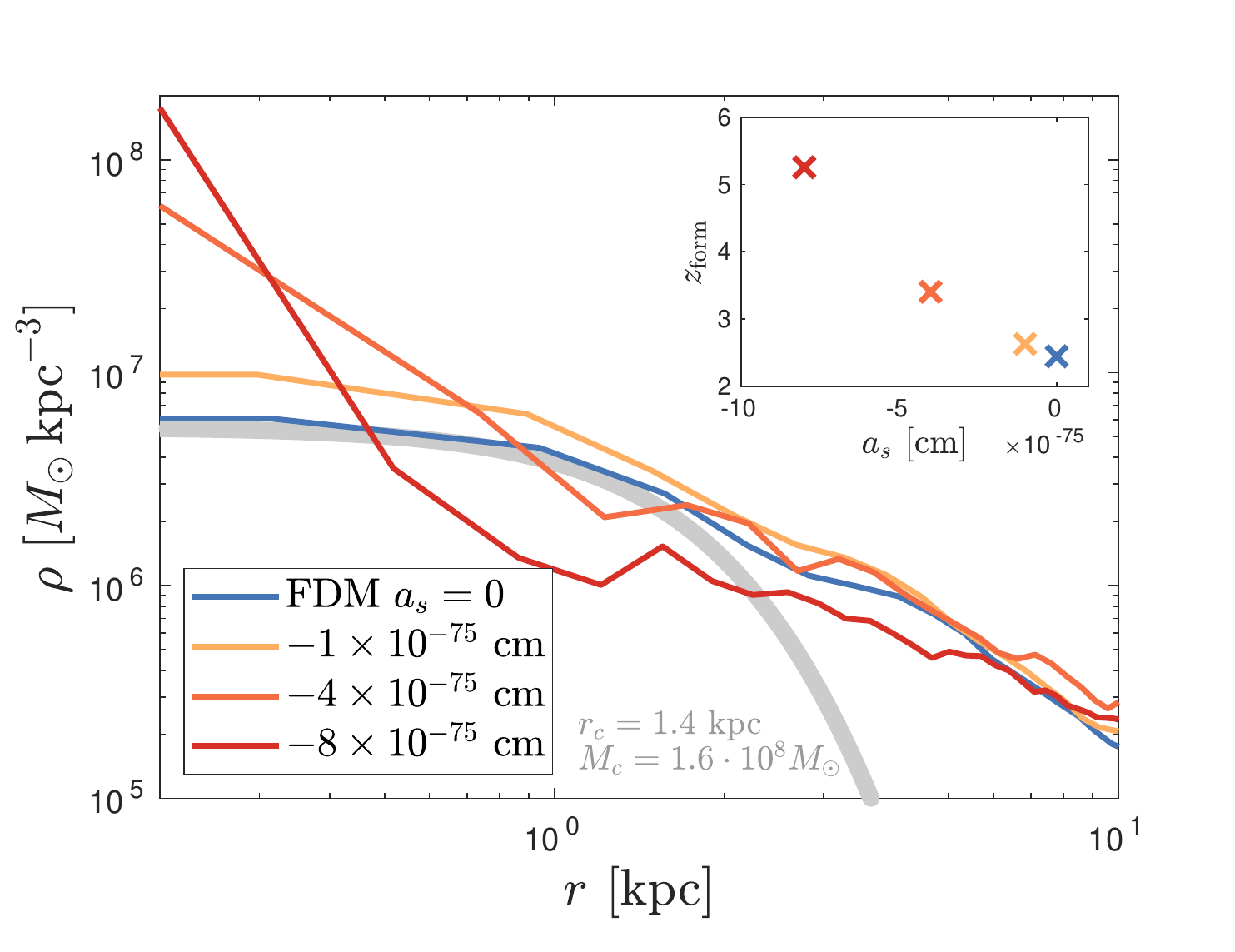}
\caption{Radial profile (physical units) of soliton at time of formation for the FDM simulations of various SI strengths. Note the presence of a \textit{dense} soliton at the two highest SI strengths. For reference, a dilute soliton profile of radius 1.4 kpc is shown (thick grey line). The inset shows the redshift of formation of the solitons as a function of the SI strength.}
\label{fig:profile}
\end{figure}

\section{Numerical Simulations}
\label{sec:sim}

In this work, we consider 5 dark matter-only simulations:
a reference CDM setup, 
a 0-self interaction FDM run, 
and FDM runs with SI characterized by a scattering length $a_s = -\{1,4,8\} \times 10^{-75}~{\rm cm}$.
Table~\ref{tab:sims}
summarizes the simulation parameters and setup, as well 
as calculates some relevant instability scales and masses described throughout the text.
The axion mass is fixed to $m=10^{-22}~{\rm eV}$.
The FDM runs have a resolution of $1024^3$ and
a box size of $L_{\rm box}=1.5~h^{-1}{\rm Mpc}$.
The box size is limited because we use a spectral method
\citep{2020MNRAS.494.2027M} to evolve the wavefunction on a uniform grid and accurately
resolve small-scale features and interference patterns that arise in solving Schr\"odinger-type systems. 
The numerical method is implemented as a module in the {\sc Arepo} code \citep{2010MNRAS.401..791S}, which is a state-of-the-art high-performance cosmological code for dark matter and baryonic simulations.
The CDM simulation is performed using the $N$-body technique with a resolution of $512^3$. Our simulations use cosmological parameters of 
$\Omega_{\rm m}= 0.3089$, $\Omega_{\rm \Lambda}= 0.6911$, $\Omega_{\rm b}= 0.0486$, $h=0.6774$ consistent with the \textit{Planck} observations of temperature and polarization anisotropies of the comic microwave background \citep{2016A&A...594A..13P}.

Initial conditions are created as a random realization of a Gaussian field,
with initial radial 1D power spectrum at redshift $z=127$ calculated by {\sc AxionCAMB} \citep{2015PhRvD..91j3512H,2000ApJ...538..473L}. All simulations are generated with the same initial random seed for phases and amplitudes, allowing for direct comparison of structures across the simulations.
In contrast to CDM, which is a scale-free theory where  dark matter structure exists on all physical scales, in FDM there is a cutoff in the dark matter power above a wavenumber
\citep{2000PhRvL..85.1158H,2017PhRvD..95d3541H}:
\begin{equation}
\label{eqn:khalf}
k_{1/2} \simeq 4.5\times  \left( \frac{m}{10^{-22}~{\rm eV}} \right)^{4/9}
{\rm Mpc}^{-1}.
\end{equation}
The simulations are run down to a redshift of $z=2$, after which the uniform resolution is insufficient to resolve small-scale structures.

The SI strengths in our simulations are set to be stronger than the fiducial value that would predict the natural abundance of dark matter via Eqn.~\ref{eqn:abundance} given our choice of axion particle mass $m=10^{-22}~{\rm eV}$, i.e.,  
$f\lesssim 10^{17}~{\rm GeV}$. 
This choice was made for a few reasons:
(1) \cite{2018PhRvD..97b3529D} estimate that
large-scale structure is impacted at lower decay constants: $f\sim 10^{13}~{\rm GeV}$.
(2) Numerical limitations make stronger SI easier to resolve on cosmological scales, and we wish to numerically verify relevant instability scales.
(3) Results may be interpolated between the SI and no SI cases.
(4) The dark matter abundance (Eqn.~\ref{eqn:abundance}) does not necessarily have to hold for all axion-like particle models.

SI additionally affects the growth of perturbations in the early Universe, but can be safely neglected in the linear regime if the axion SI is $f \lesssim  3\times 10^{15}~{\rm GeV}$ \citep{2018PhRvD..97b3529D,2021PhRvD.103l3551C}.
We have neglected the effect of SI on structure formation in the linear regime:
we have used identical initial conditions for all our FDM simulations, given by {\sc AxionCAMB} which does not include SI. 
We point out that we have chosen strong SI strengths that would actually have some moderate effect on build-up of small-scale dark matter power in the linear regime prior to the epoch our simulations are started, which we have ignored. This approach makes it more straightforward to interpolate the simulations to weaker SI, whose effects are more difficult to resolve in our cosmological volume.

Since there is a cutoff of power in the initial power spectrum (Eqn.~\ref{eqn:khalf}),
linear theory predicts that dark matter halos form only down to a particular mass
\citep{2017PhRvD..95d3541H}
\begin{equation}
\label{eqn:Mhalf}
M_{1/2} \simeq 5\times 10^{10} \left( \frac{10^{-22}~{\rm eV}}{m} \right)^{4/3} M_{\odot}.
\end{equation}
Nonlinear structure formation may support less massive quantum `pressure'-supported halos (solitons) of mass \citep{2017PhRvD..95d3541H}
\begin{equation}
\label{eqn:Mmin}
M_{\rm min} \simeq 1.4\times 10^{7} \left( \frac{10^{-22}~{\rm eV}}{m} \right)^{3/2} M_{\odot},  
\end{equation}
which has indeed been verified by numerical simulations \citep{2019PhRvL.123n1301M}.

In our study, we investigate the effect of the SI instability scale (Eqn.~\ref{eqn:kI}) on nonlinear structure formation, as well as its impact on a soliton phase transition above $M_{\rm max}$ (Eqn.~\ref{eqn:max}).

\section{Dark Matter Power Spectrum}
\label{sec:ps}

Fig.~\ref{fig:main}
shows the evolving dark matter power spectrum in our five simulations at redshifts $z=127,31,15,7,3,2$.
FDM (with and without SI) shows a reduction of power compared to CDM across all redshifts, due to the initial cutoff scale $k_{1/2}$.
However, as seen in the power spectra, the inclusion of SI leads to the growth of additional small-scale power on the instability scale $k_c$ (Eqn.~\ref{eqn:kc}).
Fig.~\ref{fig:main} marks the location of the combined instability scale $k_c$, 
as well as the individual components: the Jeans instability scale $k_J$ and the SI instability scale $k_I$.
It can be seen that for our parameters, at high redshifts $z\gtrsim 20$, the instability occurs on the SI scale $k_c\simeq k_I$,
while at lower redshifts $z\lesssim 20$, the instability is set by the Jeans scale $k_c\simeq k_J$ \citep{2021PhRvD.103l3551C}.
SI becomes less important at lower redshifts, as seen also in its scaling with $a$ in Eqn.~\eqref{eqn:kI}.

Fig.~\ref{fig:main} also shows the projected dark matter density field.
CDM is strikingly different since it forms dark matter subhalos on all spatial scales down to the numerical resolution. The FDM simulations (with and without SI) have reduced structure below $k_{1/2}$ and resemble each other more closely. However,
the inclusion of SI has slightly accelerated structure formation, which has made filaments thicker and voids less dense at the $<10$~per~cent level.
For future work, it would be of interest to study with baryonic simulations how the change in filament potentials affects star formation.
It would also be of interest to study stacked void profiles in larger scale simulations to see how they differ between CDM/FDM/WDM.

\section{Solitons}
\label{sec:soliton}

We have previously demonstrated in \cite{2019PhRvL.123n1301M} via direct numerical simulation that the first structures that form in FDM are filamentary and undergo an instability to form solitons with mass as low as $M_{\rm min}$ (Eqn.~\ref{eqn:Mmin}), which is below the cutoff scale predicted from linear theory, $M_{1/2}$ (Eqn.~\ref{eqn:Mhalf}).
We observe a similar situation in our FDM simulation, 
where we form an $M=1.6\times 10^8~M_\odot$ soliton at redshift $z=2.2$, 
which can be fit by the analytic soliton profile given by Eqn.~\eqref{eqn:soliton}.
Fig.~\ref{fig:profile} shows the measured radial profile and the analytic model, which provides a reasonable fit to the core size and central density.
The soliton mass $M$ is greater than $M_{\rm min}=1.4\times10^{7}~M_\odot$ 
but well below $M_{1/2}=5\times10^{10}~M_\odot$.
The radial profiles are plotted in terms of physical units, rather than comoving units, as solitons are physical objects detached from cosmological expansion.

It is interesting to to observe the behavior of solitons when SI is activated in the simulations
In the weakest SI case, $a_s=-1\times10^{75}~{\rm cm}$, 
solitons above $M_{\rm max} = 1.6\times 10^8~M_\odot$
are analytically expected to go unstable.
The soliton is below this threshold and thus maintains its cored shape (Fig.~\ref{fig:profile}) and is just slightly more compact and centrally concentrated due to the impact of the attractive SI.
For simulations with stronger SI, the soliton mass $M$ is now above the critical stable mass: $M > M_{\rm max}$.
Collapse is seen here, and the radial profiles are cuspy (Fig.~\ref{fig:profile}).
That is, the soliton has phase transitioned from a dilute to a dense state \citep{2018PhRvD..98b3009C}.
The simulation lacks the spatial resolution to fully resolve the final compact object with central 
density given by Eqn.~\eqref{eqn:compact}, 
which would be parsec-sized.
The critical transition from dilute to dense solitons is also further corroborated with idealized simulations of a single quasi-stationary halo in Appendix~\ref{apx:ideal}.

Fig.~\ref{fig:profile} also demonstrates that SI also leads to \textit{earlier} formation of solitons.
The redshift $z=2.2$ soliton in the no SI case forms before $z=5$ in the strongest SI simulation.
The formation of the soliton is defined as either the point in time that the filament forms an overdense $\sim$kpc core that can be approximated by the analytic soliton model, or forms a compact cusp ($<$kpc).

Finally, SI leads to the formation of \textit{additional} solitons.
The no SI FDM simulation forms just a single soliton in the filament by $z=2$ in our $1.5~h^{-1}~{\rm Mpc}$ box.
However, as indicated in Fig.~\ref{fig:main} by arrows, SI can cause constructive interference overdensities to collapse into dense solitons.
This handful of additional dense solitons are difficult to resolve due to our limited spatial resolution.

\section{Concluding Remarks}
\label{sec:conc}

We have investigated ultra-light $m=10^{-22}~{\rm eV}$ FDM simulations with an attractive SI added, 
in order to explore the effect of the axion decay constant $f$ on cosmic structure formation.
We found that an axion decay constant of $f\lesssim 10^{14}~{\rm GeV}$ leads to a noticeable increase in small-scale power.
This finding is consistent with analytic expectations for the instability scale due to attractive SI \citep{2018PhRvD..97b3529D}.
SI also leads to the formation of \textit{dense} rather than \textit{dilute} solitons above a critical mass threshold;
and thus, the prediction of the FDM model with SI is that the Universe would be populated with `bosenova' that result from cosmological initial conditions \citep{2017PhRvL.118a1301L}.
Our simulations also show that increased SI leads to the formation of additional solitons in cosmic filaments where interference patterns can cause over-densities that may be unstable under the SI.
Given the above, our work highlights the important 
changes to the model predictions of FDM
if the boson is associated with an axion and the self-coupling is taken into account.

Our work has investigated a relatively low axion decay constant ($f=5.0\times 10^{13}$--$1.4\times10^{14}~{\rm GeV}$), 
where the effects of SI are more noticeable.
Such a low value would need a physical motivation beyond the simplest models.
For a value of $f\simeq 10^{17}~{\rm GeV}$,
which is the fiducial value that predicts the total dark matter abundance in the simplest models, the attractive SI
would not have a significant impact on the structure of cosmic filaments.
The critical mass for soliton collapse would also be significantly larger: $M_{\rm max}\simeq 10^{11}~M_\odot$, which would not be cosmologically relevant to alter soliton core shapes, given the soliton core -- halo mass relation \citep{2014PhRvL.113z1302S,2021PhRvD.103l3551C}.
Hence, cosmological structure can place useful constraints on $m$ and $f$ simultaneously, which we leave for upcoming future work.
Qualitatively,
the inclusion of attractive SI goes in the right direction of 
solving the
\textit{Catch-22}
problem \citep{2020MNRAS.492.5721D} that FDM currently faces:
namely that a low particle mass $m$ is needed to predict large, low-density cores, but that erases too much structure in the high-redshift Lyman-$\alpha$ forest -- which may be recovered to an extent with SI
without the need to invoke baryonic feedback physics.

\section*{Acknowledgments}
Support (PM) for this work was provided by NASA through Einstein Postdoctoral Fellowship grant number PF7-180164 awarded by the \textit{Chandra} X-ray Center, which is operated by the Smithsonian Astrophysical Observatory for NASA under contract NAS8-03060. AF is supported by a Royal Society University Research Fellowship. MV acknowledges support through NASA ATP 19-ATP19-0019, 19-ATP19-0020, 19-ATP19-0167, and NSF grants AST-1814053, AST-1814259, AST-1909831, AST-2007355 and AST-2107724.
MBK acknowledges support from NSF CAREER award AST-1752913, NSF grants AST-1910346 and AST-2108962, NASA grant 80NSSC22K0827, and HST-AR-15809, HST-GO-15658, HST-GO-15901, HST-GO-15902, HST-AR-16159, and HST-GO-16226 from the Space Telescope Science Institute, which is operated by AURA, Inc., under NASA contract NAS5-26555. MA is supported by a NASA grant 80NSSC20K0518. SB is supported by the UK Research and Innovation (UKRI) Future Leaders Fellowship [grant number MR/V023381/1]. TD acknowledges support from the Isaac Newton Studentship and the Science and Technology Facilities Council under grant number ST/V50659X/1. This work was performed using the Cambridge Service for Data Driven Discovery (CSD3), part of which is operated by the University of Cambridge Research Computing on behalf of the STFC DiRAC HPC Facility (\url{www.dirac.ac.uk}). The DiRAC component of CSD3 was funded by BEIS capital funding via STFC capital grants ST/P002307/1 and ST/R002452/1 and STFC operations grant ST/R00689X/1. DiRAC is part of the National e-Infrastructure.
This material is based upon work supported by the National Science Foundation under Grant DGE-2108962.
This work was performed under the auspices of the U.S. Department of Energy by Lawrence Livermore National Laboratory under contract DE-AC52-07NA27344. Lawrence Livermore National Security, LLC.

\section*{Data Availability Statement}

The data underlying this article will be shared on reasonable request to the corresponding author.

\bibliography{mybib}{}

\begin{thebibliography}{}
\expandafter\ifx\csname natexlab\endcsname\relax\def\natexlab#1{#1}\fi
\providecommand{\url}[1]{\href{#1}{#1}}
\providecommand{\dodoi}[1]{doi:~\href{http://doi.org/#1}{\nolinkurl{#1}}}
\providecommand{\doeprint}[1]{\href{http://ascl.net/#1}{\nolinkurl{http://ascl.net/#1}}}
\providecommand{\doarXiv}[1]{\href{https://arxiv.org/abs/#1}{\nolinkurl{https://arxiv.org/abs/#1}}}

\bibitem[{Amin \& Mocz(2019)}]{Amin:2019ums}
Amin, M.~A., \& Mocz, P. 2019, Phys. Rev. D, 100, 063507,
  \dodoi{10.1103/PhysRevD.100.063507}

\bibitem[{{Braaten} {et~al.}(2016){Braaten}, {Mohapatra}, \&
  {Zhang}}]{2016PhRvL.117l1801B}
{Braaten}, E., {Mohapatra}, A., \& {Zhang}, H. 2016, \prl, 117, 121801,
  \dodoi{10.1103/PhysRevLett.117.121801}

\bibitem[{{Braaten} \& {Zhang}(2019)}]{2019RvMP...91d1002B}
{Braaten}, E., \& {Zhang}, H. 2019, Reviews of Modern Physics, 91, 041002,
  \dodoi{10.1103/RevModPhys.91.041002}

\bibitem[{{Burkert}(2020)}]{2020ApJ...904..161B}
{Burkert}, A. 2020, \apj, 904, 161, \dodoi{10.3847/1538-4357/abb242}

\bibitem[{{Chavanis}(2011)}]{2011PhRvD..84d3531C}
{Chavanis}, P.-H. 2011, \prd, 84, 043531, \dodoi{10.1103/PhysRevD.84.043531}

\bibitem[{{Chavanis}(2012)}]{2012A&A...537A.127C}
---. 2012, \aap, 537, A127, \dodoi{10.1051/0004-6361/201116905}

\bibitem[{{Chavanis}(2016)}]{2016PhRvD..94h3007C}
---. 2016, \prd, 94, 083007, \dodoi{10.1103/PhysRevD.94.083007}

\bibitem[{{Chavanis}(2018)}]{2018PhRvD..98b3009C}
---. 2018, \prd, 98, 023009, \dodoi{10.1103/PhysRevD.98.023009}

\bibitem[{{Chavanis}(2019)}]{2019PhRvD.100h3022C}
---. 2019, \prd, 100, 083022, \dodoi{10.1103/PhysRevD.100.083022}

\bibitem[{{Chavanis}(2021)}]{2021PhRvD.103l3551C}
---. 2021, \prd, 103, 123551, \dodoi{10.1103/PhysRevD.103.123551}

\bibitem[{{Chavanis} \& {Delfini}(2011)}]{2011PhRvD..84d3532C}
{Chavanis}, P.-H., \& {Delfini}, L. 2011, \prd, 84, 043532,
  \dodoi{10.1103/PhysRevD.84.043532}

\bibitem[{{Chen} {et~al.}(2021){Chen}, {Du}, {Lentz}, {Marsh}, \&
  {Niemeyer}}]{2021PhRvD.104h3022C}
{Chen}, J., {Du}, X., {Lentz}, E.~W., {Marsh}, D. J.~E., \& {Niemeyer}, J.~C.
  2021, \prd, 104, 083022, \dodoi{10.1103/PhysRevD.104.083022}

\bibitem[{Dalal \& Kravtsov(2022)}]{dalal2022not}
Dalal, N., \& Kravtsov, A. 2022, arXiv preprint arXiv:2203.05750

\bibitem[{{Davies} \& {Mocz}(2020)}]{2020MNRAS.492.5721D}
{Davies}, E.~Y., \& {Mocz}, P. 2020, \mnras, 492, 5721,
  \dodoi{10.1093/mnras/staa202}

\bibitem[{{Desjacques} {et~al.}(2018){Desjacques}, {Kehagias}, \&
  {Riotto}}]{2018PhRvD..97b3529D}
{Desjacques}, V., {Kehagias}, A., \& {Riotto}, A. 2018, \prd, 97, 023529,
  \dodoi{10.1103/PhysRevD.97.023529}

\bibitem[{{Di Vecchia} \& {Veneziano}(1980)}]{1980NuPhB.171..253D}
{Di Vecchia}, P., \& {Veneziano}, G. 1980, Nuclear Physics B, 171, 253,
  \dodoi{10.1016/0550-3213(80)90370-3}

\bibitem[{{Du} {et~al.}(2018){Du}, {Schwabe}, {Niemeyer}, \&
  {B{\"u}rger}}]{2018PhRvD..97f3507D}
{Du}, X., {Schwabe}, B., {Niemeyer}, J.~C., \& {B{\"u}rger}, D. 2018, \prd, 97,
  063507, \dodoi{10.1103/PhysRevD.97.063507}

\bibitem[{{Eby} {et~al.}(2019){Eby}, {Leembruggen}, {Street}, {Suranyi}, \&
  {Wijewardhana}}]{2019PhRvD.100f3002E}
{Eby}, J., {Leembruggen}, M., {Street}, L., {Suranyi}, P., \& {Wijewardhana},
  L.~C.~R. 2019, \prd, 100, 063002, \dodoi{10.1103/PhysRevD.100.063002}

\bibitem[{{Eby} {et~al.}(2016){Eby}, {Leembruggen}, {Suranyi}, \&
  {Wijewardhana}}]{2016JHEP...12..066E}
{Eby}, J., {Leembruggen}, M., {Suranyi}, P., \& {Wijewardhana}, L.~C.~R. 2016,
  Journal of High Energy Physics, 2016, 66, \dodoi{10.1007/JHEP12(2016)066}

\bibitem[{{Glennon} \& {Prescod-Weinstein}(2021)}]{2021PhRvD.104h3532G}
{Glennon}, N., \& {Prescod-Weinstein}, C. 2021, \prd, 104, 083532,
  \dodoi{10.1103/PhysRevD.104.083532}

\bibitem[{{Guzm{\'a}n} \& {Ure{\~n}a-L{\'o}pez}(2003)}]{2003PhRvD..68b4023G}
{Guzm{\'a}n}, F., \& {Ure{\~n}a-L{\'o}pez}, L. 2003, \prd, 68, 024023,
  \dodoi{10.1103/PhysRevD.68.024023}

\bibitem[{{Helfer} {et~al.}(2017){Helfer}, {Marsh}, {Clough}, {Fairbairn},
  {Lim}, \& {Becerril}}]{2017JCAP...03..055H}
{Helfer}, T., {Marsh}, D. J.~E., {Clough}, K., {et~al.} 2017, \jcap, 2017, 055,
  \dodoi{10.1088/1475-7516/2017/03/055}

\bibitem[{{Hlozek} {et~al.}(2015){Hlozek}, {Grin}, {Marsh}, \&
  {Ferreira}}]{2015PhRvD..91j3512H}
{Hlozek}, R., {Grin}, D., {Marsh}, D. J.~E., \& {Ferreira}, P.~G. 2015, \prd,
  91, 103512, \dodoi{10.1103/PhysRevD.91.103512}

\bibitem[{{Hu} {et~al.}(2000){Hu}, {Barkana}, \&
  {Gruzinov}}]{2000PhRvL..85.1158H}
{Hu}, W., {Barkana}, R., \& {Gruzinov}, A. 2000, \prl, 85, 1158,
  \dodoi{10.1103/PhysRevLett.85.1158}

\bibitem[{{Hui}(2021)}]{2021arXiv210111735H}
{Hui}, L. 2021, arXiv e-prints, arXiv:2101.11735.
\newblock \doarXiv{2101.11735}

\bibitem[{{Hui} {et~al.}(2017){Hui}, {Ostriker}, {Tremaine}, \&
  {Witten}}]{2017PhRvD..95d3541H}
{Hui}, L., {Ostriker}, J.~P., {Tremaine}, S., \& {Witten}, E. 2017, \prd, 95,
  043541, \dodoi{10.1103/PhysRevD.95.043541}

\bibitem[{{Ir{\v{s}}i{\v{c}}} {et~al.}(2017){Ir{\v{s}}i{\v{c}}}, {Viel},
  {Haehnelt}, {Bolton}, \& {Becker}}]{2017PhRvL.119c1302I}
{Ir{\v{s}}i{\v{c}}}, V., {Viel}, M., {Haehnelt}, M.~G., {Bolton}, J.~S., \&
  {Becker}, G.~D. 2017, \prl, 119, 031302,
  \dodoi{10.1103/PhysRevLett.119.031302}

\bibitem[{{Khlopov} {et~al.}(1985){Khlopov}, {Malomed}, \&
  {Zeldovich}}]{1985MNRAS.215..575K}
{Khlopov}, M.~I., {Malomed}, B.~A., \& {Zeldovich}, I.~B. 1985, \mnras, 215,
  575, \dodoi{10.1093/mnras/215.4.575}

\bibitem[{{Lagu{\"e}} {et~al.}(2020){Lagu{\"e}}, {Bond}, {Hlo{\v{z}}ek},
  {Marsh}, \& {S{\"o}ding}}]{2020arXiv200408482L}
{Lagu{\"e}}, A., {Bond}, J.~R., {Hlo{\v{z}}ek}, R., {Marsh}, D. J.~E., \&
  {S{\"o}ding}, L. 2020, arXiv e-prints, arXiv:2004.08482.
\newblock \doarXiv{2004.08482}

\bibitem[{{Levkov} {et~al.}(2017){Levkov}, {Panin}, \&
  {Tkachev}}]{2017PhRvL.118a1301L}
{Levkov}, D.~G., {Panin}, A.~G., \& {Tkachev}, I.~I. 2017, \prl, 118, 011301,
  \dodoi{10.1103/PhysRevLett.118.011301}

\bibitem[{{Lewis} {et~al.}(2000){Lewis}, {Challinor}, \&
  {Lasenby}}]{2000ApJ...538..473L}
{Lewis}, A., {Challinor}, A., \& {Lasenby}, A. 2000, \apj, 538, 473,
  \dodoi{10.1086/309179}

\bibitem[{{Li} {et~al.}(2021){Li}, {Hui}, \& {Yavetz}}]{2021PhRvD.103b3508L}
{Li}, X., {Hui}, L., \& {Yavetz}, T.~D. 2021, \prd, 103, 023508,
  \dodoi{10.1103/PhysRevD.103.023508}

\bibitem[{{Marsh}(2016)}]{2016PhR...643....1M}
{Marsh}, D. J.~E. 2016, \physrep, 643, 1, \dodoi{10.1016/j.physrep.2016.06.005}

\bibitem[{{Marsh} \& {Pop}(2015)}]{2015MNRAS.451.2479M}
{Marsh}, D. J.~E., \& {Pop}, A.-R. 2015, \mnras, 451, 2479,
  \dodoi{10.1093/mnras/stv1050}

\bibitem[{{May} \& {Springel}(2021)}]{2021arXiv210101828M}
{May}, S., \& {Springel}, V. 2021, arXiv e-prints, arXiv:2101.01828.
\newblock \doarXiv{2101.01828}

\bibitem[{{Mocz} {et~al.}(2018){Mocz}, {Lancaster}, {Fialkov}, {Becerra}, \&
  {Chavanis}}]{2018PhRvD..97h3519M}
{Mocz}, P., {Lancaster}, L., {Fialkov}, A., {Becerra}, F., \& {Chavanis}, P.-H.
  2018, \prd, 97, 083519, \dodoi{10.1103/PhysRevD.97.083519}

\bibitem[{{Mocz} {et~al.}(2017){Mocz}, {Vogelsberger}, {Robles}, {Zavala},
  {Boylan-Kolchin}, {Fialkov}, \& {Hernquist}}]{2017MNRAS.471.4559M}
{Mocz}, P., {Vogelsberger}, M., {Robles}, V.~H., {et~al.} 2017, \mnras, 471,
  4559, \dodoi{10.1093/mnras/stx1887}

\bibitem[{{Mocz} {et~al.}(2019){Mocz}, {Fialkov}, {Vogelsberger}, {Becerra},
  {Amin}, {Bose}, {Boylan-Kolchin}, {Chavanis}, {Hernquist}, {Lancaster},
  {Marinacci}, {Robles}, \& {Zavala}}]{2019PhRvL.123n1301M}
{Mocz}, P., {Fialkov}, A., {Vogelsberger}, M., {et~al.} 2019, \prl, 123,
  141301, \dodoi{10.1103/PhysRevLett.123.141301}

\bibitem[{{Mocz} {et~al.}(2020){Mocz}, {Fialkov}, {Vogelsberger}, {Becerra},
  {Shen}, {Robles}, {Amin}, {Zavala}, {Boylan-Kolchin}, {Bose}, {Marinacci},
  {Chavanis}, {Lancaster}, \& {Hernquist}}]{2020MNRAS.494.2027M}
---. 2020, \mnras, 494, 2027, \dodoi{10.1093/mnras/staa738}

\bibitem[{{Niemeyer}(2020)}]{2020PrPNP.11303787N}
{Niemeyer}, J.~C. 2020, Progress in Particle and Nuclear Physics, 113, 103787,
  \dodoi{10.1016/j.ppnp.2020.103787}

\bibitem[{{Nori} \& {Baldi}(2021)}]{2021MNRAS.501.1539N}
{Nori}, M., \& {Baldi}, M. 2021, \mnras, 501, 1539,
  \dodoi{10.1093/mnras/staa3772}

\bibitem[{{Nori} {et~al.}(2019){Nori}, {Murgia}, {Ir{\v{s}}i{\v{c}}}, {Baldi},
  \& {Viel}}]{2019MNRAS.482.3227N}
{Nori}, M., {Murgia}, R., {Ir{\v{s}}i{\v{c}}}, V., {Baldi}, M., \& {Viel}, M.
  2019, \mnras, 482, 3227, \dodoi{10.1093/mnras/sty2888}

\bibitem[{{Peccei} \& {Quinn}(1977)}]{1977PhRvL..38.1440P}
{Peccei}, R.~D., \& {Quinn}, H.~R. 1977, \prl, 38, 1440,
  \dodoi{10.1103/PhysRevLett.38.1440}

\bibitem[{{Planck Collaboration} {et~al.}(2016){Planck Collaboration}, {Ade},
  {Aghanim}, {Arnaud}, {Ashdown}, {Aumont}, {Baccigalupi}, {Banday},
  {Barreiro}, {Bartlett}, {Bartolo}, {Battaner}, {Battye}, {Benabed},
  {Beno{\^\i}t}, {Benoit-L{\'e}vy}, {Bernard}, {Bersanelli}, {Bielewicz},
  {Bock}, {Bonaldi}, {Bonavera}, {Bond}, {Borrill}, {Bouchet}, {Boulanger},
  {Bucher}, {Burigana}, {Butler}, {Calabrese}, {Cardoso}, {Catalano},
  {Challinor}, {Chamballu}, {Chary}, {Chiang}, {Chluba}, {Christensen},
  {Church}, {Clements}, {Colombi}, {Colombo}, {Combet}, {Coulais}, {Crill},
  {Curto}, {Cuttaia}, {Danese}, {Davies}, {Davis}, {de Bernardis}, {de Rosa},
  {de Zotti}, {Delabrouille}, {D{\'e}sert}, {Di Valentino}, {Dickinson},
  {Diego}, {Dolag}, {Dole}, {Donzelli}, {Dor{\'e}}, {Douspis}, {Ducout},
  {Dunkley}, {Dupac}, {Efstathiou}, {Elsner}, {En{\ss}lin}, {Eriksen},
  {Farhang}, {Fergusson}, {Finelli}, {Forni}, {Frailis}, {Fraisse},
  {Franceschi}, {Frejsel}, {Galeotta}, {Galli}, {Ganga}, {Gauthier}, {Gerbino},
  {Ghosh}, {Giard}, {Giraud-H{\'e}raud}, {Giusarma}, {Gjerl{\o}w},
  {Gonz{\'a}lez-Nuevo}, {G{\'o}rski}, {Gratton}, {Gregorio}, {Gruppuso},
  {Gudmundsson}, {Hamann}, {Hansen}, {Hanson}, {Harrison}, {Helou},
  {Henrot-Versill{\'e}}, {Hern{\'a}ndez-Monteagudo}, {Herranz}, {Hildebrandt},
  {Hivon}, {Hobson}, {Holmes}, {Hornstrup}, {Hovest}, {Huang}, {Huffenberger},
  {Hurier}, {Jaffe}, {Jaffe}, {Jones}, {Juvela}, {Keih{\"a}nen}, {Keskitalo},
  {Kisner}, {Kneissl}, {Knoche}, {Knox}, {Kunz}, {Kurki-Suonio}, {Lagache},
  {L{\"a}hteenm{\"a}ki}, {Lamarre}, {Lasenby}, {Lattanzi}, {Lawrence}, {Leahy},
  {Leonardi}, {Lesgourgues}, {Levrier}, {Lewis}, {Liguori}, {Lilje},
  {Linden-V{\o}rnle}, {L{\'o}pez-Caniego}, {Lubin}, {Mac{\'\i}as-P{\'e}rez},
  {Maggio}, {Maino}, {Mandolesi}, {Mangilli}, {Marchini}, {Maris}, {Martin},
  {Martinelli}, {Mart{\'\i}nez-Gonz{\'a}lez}, {Masi}, {Matarrese}, {McGehee},
  {Meinhold}, {Melchiorri}, {Melin}, {Mendes}, {Mennella}, {Migliaccio},
  {Millea}, {Mitra}, {Miville-Desch{\^e}nes}, {Moneti}, {Montier}, {Morgante},
  {Mortlock}, {Moss}, {Munshi}, {Murphy}, {Naselsky}, {Nati}, {Natoli},
  {Netterfield}, {N{\o}rgaard-Nielsen}, {Noviello}, {Novikov}, {Novikov},
  {Oxborrow}, {Paci}, {Pagano}, {Pajot}, {Paladini}, {Paoletti}, {Partridge},
  {Pasian}, {Patanchon}, {Pearson}, {Perdereau}, {Perotto}, {Perrotta},
  {Pettorino}, {Piacentini}, {Piat}, {Pierpaoli}, {Pietrobon}, {Plaszczynski},
  {Pointecouteau}, {Polenta}, {Popa}, {Pratt}, {Pr{\'e}zeau}, {Prunet},
  {Puget}, {Rachen}, {Reach}, {Rebolo}, {Reinecke}, {Remazeilles}, {Renault},
  {Renzi}, {Ristorcelli}, {Rocha}, {Rosset}, {Rossetti}, {Roudier},
  {Rouill{\'e} d'Orfeuil}, {Rowan-Robinson}, {Rubi{\~n}o-Mart{\'\i}n},
  {Rusholme}, {Said}, {Salvatelli}, {Salvati}, {Sandri}, {Santos},
  {Savelainen}, {Savini}, {Scott}, {Seiffert}, {Serra}, {Shellard}, {Spencer},
  {Spinelli}, {Stolyarov}, {Stompor}, {Sudiwala}, {Sunyaev}, {Sutton},
  {Suur-Uski}, {Sygnet}, {Tauber}, {Terenzi}, {Toffolatti}, {Tomasi},
  {Tristram}, {Trombetti}, {Tucci}, {Tuovinen}, {T{\"u}rler}, {Umana},
  {Valenziano}, {Valiviita}, {Van Tent}, {Vielva}, {Villa}, {Wade}, {Wandelt},
  {Wehus}, {White}, {White}, {Wilkinson}, {Yvon}, {Zacchei}, \&
  {Zonca}}]{2016A&A...594A..13P}
{Planck Collaboration}, {Ade}, P.~A.~R., {Aghanim}, N., {et~al.} 2016, \aap,
  594, A13, \dodoi{10.1051/0004-6361/201525830}

\bibitem[{{Safarzadeh} \& {Spergel}(2020)}]{2020ApJ...893...21S}
{Safarzadeh}, M., \& {Spergel}, D.~N. 2020, \apj, 893, 21,
  \dodoi{10.3847/1538-4357/ab7db2}

\bibitem[{{Schive} {et~al.}(2014{\natexlab{a}}){Schive}, {Chiueh}, \&
  {Broadhurst}}]{2014NatPh..10..496S}
{Schive}, H.-Y., {Chiueh}, T., \& {Broadhurst}, T. 2014{\natexlab{a}}, Nature
  Physics, 10, 496, \dodoi{10.1038/nphys2996}

\bibitem[{{Schive} {et~al.}(2014{\natexlab{b}}){Schive}, {Liao}, {Woo}, {Wong},
  {Chiueh}, {Broadhurst}, \& {Hwang}}]{2014PhRvL.113z1302S}
{Schive}, H.-Y., {Liao}, M.-H., {Woo}, T.-P., {et~al.} 2014{\natexlab{b}},
  \prl, 113, 261302, \dodoi{10.1103/PhysRevLett.113.261302}

\bibitem[{{Schutz}(2020)}]{2020PhRvD.101l3026S}
{Schutz}, K. 2020, \prd, 101, 123026, \dodoi{10.1103/PhysRevD.101.123026}

\bibitem[{{Schwabe} {et~al.}(2020){Schwabe}, {Gosenca}, {Behrens}, {Niemeyer},
  \& {Easther}}]{2020PhRvD.102h3518S}
{Schwabe}, B., {Gosenca}, M., {Behrens}, C., {Niemeyer}, J.~C., \& {Easther},
  R. 2020, \prd, 102, 083518, \dodoi{10.1103/PhysRevD.102.083518}

\bibitem[{{Springel}(2010)}]{2010MNRAS.401..791S}
{Springel}, V. 2010, \mnras, 401, 791, \dodoi{10.1111/j.1365-2966.2009.15715.x}

\bibitem[{{Vakhitov} \& {Kolokolov}(1973)}]{1973R&QE...16..783V}
{Vakhitov}, N.~G., \& {Kolokolov}, A.~A. 1973, Radiophysics and Quantum
  Electronics, 16, 783, \dodoi{10.1007/BF01031343}

\bibitem[{{Veltmaat} {et~al.}(2020){Veltmaat}, {Schwabe}, \&
  {Niemeyer}}]{2020PhRvD.101h3518V}
{Veltmaat}, J., {Schwabe}, B., \& {Niemeyer}, J.~C. 2020, \prd, 101, 083518,
  \dodoi{10.1103/PhysRevD.101.083518}

\bibitem[{{Visinelli} {et~al.}(2018){Visinelli}, {Baum}, {Redondo}, {Freese},
  \& {Wilczek}}]{2018PhLB..777...64V}
{Visinelli}, L., {Baum}, S., {Redondo}, J., {Freese}, K., \& {Wilczek}, F.
  2018, Physics Letters B, 777, 64, \dodoi{10.1016/j.physletb.2017.12.010}

\bibitem[{{Weinberg}(1978)}]{1978PhRvL..40..223W}
{Weinberg}, S. 1978, \prl, 40, 223, \dodoi{10.1103/PhysRevLett.40.223}

\bibitem[{{Witten}(1980)}]{1980AnPhy.128..363W}
{Witten}, E. 1980, Annals of Physics, 128, 363,
  \dodoi{10.1016/0003-4916(80)90325-5}

\end{thebibliography}

\appendix

\section{Idealized Simulations of Soliton Phase Transition}
\label{apx:ideal}

We perform additional simulations of an idealized FDM halo with SI, 
to confirm the transition from dilute to dense solitons above the critical mass $M_{\rm max}$ \citep{2018PhRvD..98b3009C}
in an idealized setting with higher effective resolution of the core.
The setup follows \cite{2017MNRAS.471.4559M}, where random solitons are merged to form a single quasi-stationary  halo. The simulation has a box size of $L=20~{\rm kpc}$, resolution $400^3$, axion mass $m=10^{-22}~{\rm eV}$, and is run for $4~{\rm Gyr}$.
In the reference case with SI  switched off ($a_s=0$), the result is a quasi-stationary  halo 
with a soliton core of mass $M=1.2\times 10^9~M_\odot$ (radius $0.2~{\rm kpc}$).
We consider additional simulation cases
with SI strengths:
$a_s = -\{ 0.5,0.9,1.4,1.5,1.6,2  \}\times10^{-77}~{\rm cm}$,
corresponding to
$f=\{ 2, 1.5, 1.2, 1.15, 1.125, 1 \}\times10^{15}~{\rm GeV}$.

\begin{figure}[ht!]
\centering
\includegraphics[width=0.6\textwidth]{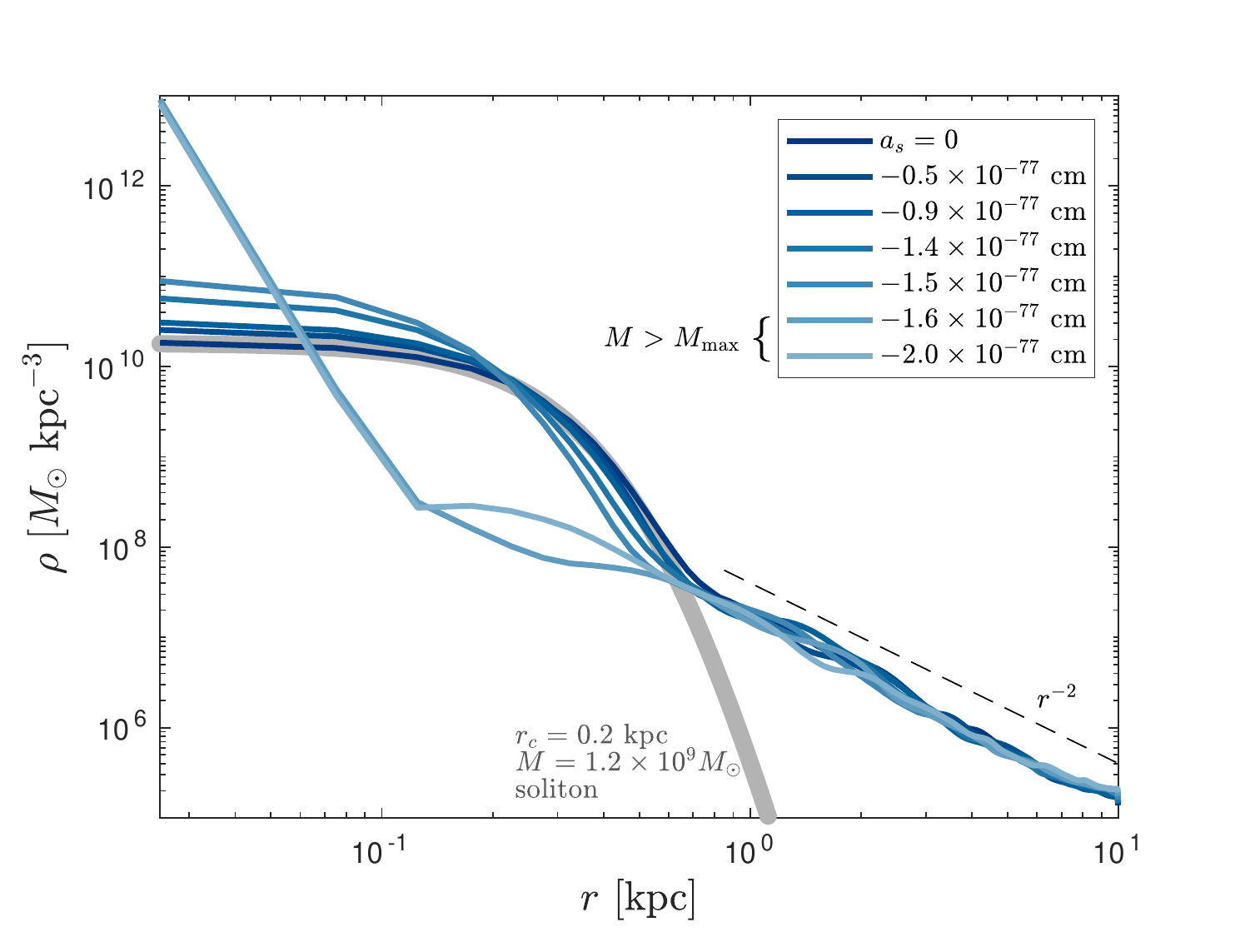}\\
\begin{tabular}{ccccccc}
\includegraphics[width=0.124\textwidth]{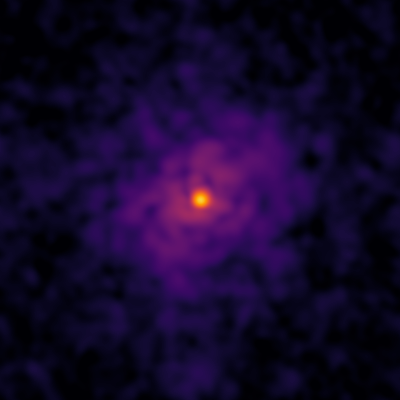} &
\includegraphics[width=0.124\textwidth]{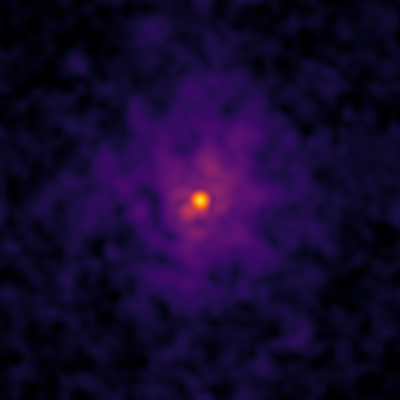} &
\includegraphics[width=0.124\textwidth]{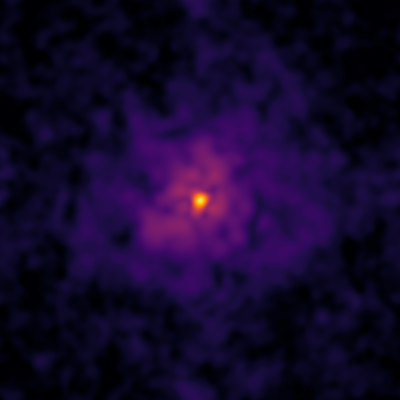} &
\includegraphics[width=0.124\textwidth]{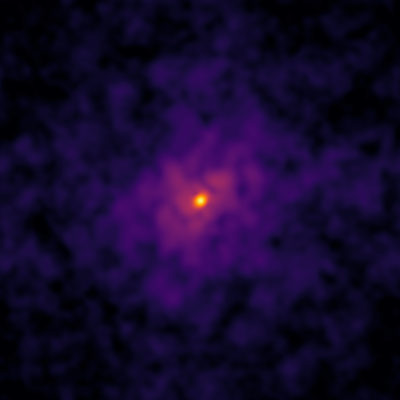} &
\includegraphics[width=0.124\textwidth]{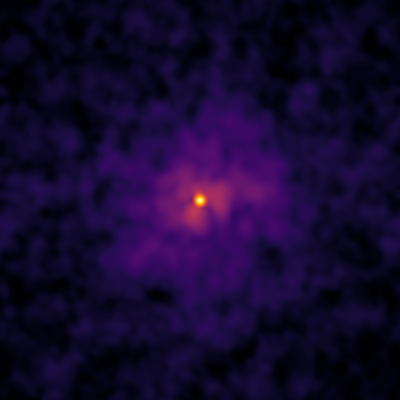} &
\includegraphics[width=0.124\textwidth]{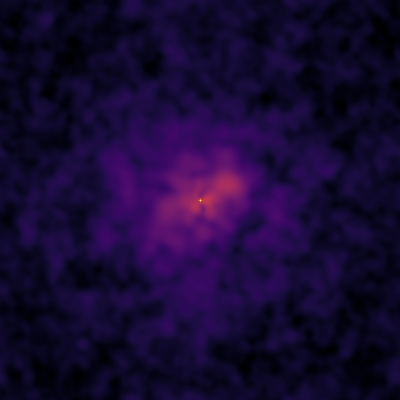} &
\includegraphics[width=0.124\textwidth]{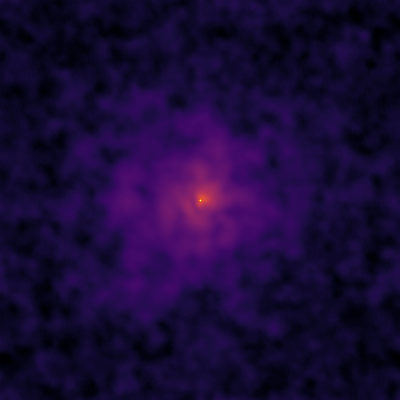} \\
 $\left(\frac{a_s}{10^{-77}~{\rm cm}}\right)=0$ &  
 $-0.5$ &  
 $-0.9$ & 
 $-1.4$ &
 $-1.5$ &
 $-1.6$ &
 $-2.0$ \\
\end{tabular}
\caption{Radial profiles and projected densities for idealized FDM halo with SI. A phase transition is observed to occur in the central soliton at large attractive SI strengths. For reference, a dilute soliton profile of radius 2 kpc is shown (thick grey line) }
\label{fig:ideal}
\end{figure}

Fig.~\ref{fig:ideal} shows the resulting radial profiles of the halo for each SI strength.
As the attractive SI strength increases, the soliton becomes more dense and compact.
The phase transition is observed when the soliton mass is $M>M_{\rm max}$, which is the case for the
two strongest SI strengths simulated.
The outer radial profile of the dark matter halo is close to an $r^{-2}$ isothermal profile, 
as analytically predicted in \cite{2019PhRvD.100h3022C}, 
and is largely unaffected by the collapse of the central soliton.
A more detailed study of idealized collapse will be presented by Painter et al. (in prep).

\end{document}